\documentclass[%
 reprint,
superscriptaddress,
 amsmath,amssymb,
 aps,
]{revtex4-2}

\usepackage{graphicx}
\usepackage{dcolumn}
\usepackage{bm}

\begin{document}

\preprint{APS/123-QED}

\title{Experimental Verification of Circumferential Bunch Length Variation and Head-Tail Exchange Affecting Microwave Instability in a Storage Ring}

\author{Jihong Bian}
\affiliation{Department of Engineering Physics, Tsinghua University, Beijing 100084, China}%

\author{Xiujie Deng}\email{Contact author: dengxiujie@mail.tsinghua.edu.cn}
\affiliation{Institute for Advanced Study, Tsinghua University, Beijing 100084, China}%

\author{Arne Hoehl}
\affiliation{Physikalisch-Technische Bundesanstalt (PTB), Abbestra$\beta$e 2-12, 10587 Berlin, Germany}%

\author{Wenhui Huang}\email{Contact author: huangwh@mail.tsinghua.edu.cn}
\affiliation{Department of Engineering Physics, Tsinghua University, Beijing 100084, China}%

\author{Arnold~Kruschinski}\email{Contact author: arnold.kruschinski@helmholtz-berlin.de}
\affiliation{Helmholtz-Zentrum Berlin (HZB), Albert-Einstein-Stra$\beta$e 15, 12489 Berlin, Germany}%

\author{Carsten Mai}%
\affiliation{Helmholtz-Zentrum Berlin (HZB), Albert-Einstein-Stra$\beta$e 15, 12489 Berlin, Germany}%

\author{Markus Ries}
\affiliation{Helmholtz-Zentrum Berlin (HZB), Albert-Einstein-Stra$\beta$e 15, 12489 Berlin, Germany}%

\author{Chuanxiang Tang}
\affiliation{Department of Engineering Physics, Tsinghua University, Beijing 100084, China}%

\date{\today}

\begin{abstract}
Classical analyses of microwave instability are built upon the longitudinal adiabatic approximation, which assumes that the bunch length remains constant around the storage ring. However, in a storage ring with small global phase slippage, the bunch length can vary around the ring and some particles can experience head-tail exchange due to the partial phase slippage and transverse-longitudinal coupling. Our theoretical study reveals that these effects can be beneficial for suppressing microwave instability. A new microwave instability threshold evaluation method has been correspondingly proposed to account for these effects. Here we present the first experimental evidence supporting our theoretical analysis. The measurements confirm that the microwave instability threshold can be increased by a factor of up to six compared to the classical prediction in our cases. Our results can also provide practical guidance for the design of extremely short bunch storage rings.
\end{abstract}

\maketitle

Accelerator-based light sources have proven indispensable across numerous research fields. Storage ring-based synchrotron radiation sources and linear accelerator (linac)-based free electron lasers (FELs) are currently two major categories of such facilities, which can deliver radiation with high repetition rate and high peak power, respectively. Recently, a novel storage ring-based light source mechanism called steady-state microbunching (SSMB) has attracted extensive research interest and made substantial progress \cite{PhysRevLett.105.154801,PhysRevSTAB.14.110702,Chao:IPAC2016-TUXB01,Tang:FLS2018-THP2WB02,PhysRevAccelBeams.23.044002,PhysRevAccelBeams.23.044001,PhysRevAccelBeams.24.090701,PhysRevAccelBeams.24.094001,PhysRevAccelBeams.26.054001,PhysRevAccelBeams.26.110701,PhysRevAccelBeams.24.114401,PhysRevAccelBeams.25.064401,TSAI2022167454,kwxj-4bff,jp4b-1mhy,PhysRevAccelBeams.27.094201,kxbb-n27x,ysdm-mzv3,3v8n-84fr,c4zc-mghl,h677-zjdg,15-20220486,deng2026steady}, especially the success of the SSMB proof-of-principle experiment \cite{deng2021experimental,kruschinski2024confirming}. This scheme aims to generate electron bunches shorter than the targeted radiation wavelength inside a storage ring to produce coherent radiation that promises both high repetition rate and high peak power, and thus holds tremendous potential for diverse applications.

To achieve ultrashort electron bunches in a storage ring, a direct strategy is to reduce the storage ring global phase slippage $\eta$, which follows the classical “zero-current” bunch length scaling law $\sigma_z \propto \sqrt{|\eta|}$, given by Sands \cite{Sands:102780}. The global phase slippage describes the energy dependence of particle’s revolution time: 
\begin{equation}
	\eta = \frac{\Delta T/T_0}{\Delta E/E_0}  = 
	\frac{1}{C_0} \oint \left( \frac{D_x(s)}{\rho(s)} - \frac{1}{\gamma^2} \right) ds,
	\label{eq:1}
\end{equation}
where $T_0$ is the revolution period, $E_0=\gamma m_e c^2$ is the energy of the reference particle, $m_e$ is the electron rest mass, $\gamma$ is the Lorentz factor, $C_0$ is the circumference of the ring, $D_x$ is the horizontal dispersion, $\rho$ is the bending radius of the trajectory, and $s$ is the path length along the reference orbit. However, when the global phase slippage becomes extremely small, the effect of the partial phase slippage can significantly modify the single-particle longitudinal dynamics compared with the classical analysis. In this regime, the equilibrium energy spread diverges due to the quantum diffusion, which has been verified both theoretically and experimentally \cite{PhysRevAccelBeams.24.094001,PhysRevAccelBeams.26.054001}. Another novel dynamic feature is that the bunch length is no longer a constant over one turn along the ring but can vary substantially. Further considering the effect of transverse-longitudinal coupling, the more accurate formula of bunch length is
\begin{equation}
	\sigma_z = \sqrt{\epsilon_z \beta_z + \epsilon_x \mathcal{H}_x},
	\label{eq:2}
\end{equation}
where $\epsilon_z$ and $\epsilon_x$ are the longitudinal and horizontal emittance, $\beta_z$ is the longitudinal beta function, and $\mathcal{H}_x = \gamma_x {D_x}^2 + 2\alpha_x D_x D_x' + \beta_x {D_x'}^2$ is the horizontal chromatic function.

Collective instabilities fundamentally limit the achievable beam current in storage rings. Bane \textit{et al.} have given a threshold formula for coherent synchrotron radiation (CSR) induced microwave instability (MWI), which is the dominant effect in short bunch storage rings \cite{PhysRevSTAB.13.104402}. Experimental measurements performed at the Karlsruhe Research Accelerator (KARA) exhibit good agreement with this theoretical formula \cite{PhysRevAccelBeams.22.020701}. However, this classical formula is derived based on the longitudinal adiabatic approximation, which assumes that the bunch length remains constant around the ring. Our further theoretical studies reveal that when the global phase slippage is extremely small, the bunch length variation around the ring and head-tail exchange of some particles caused by the partial phase slippage and transverse-longitudinal coupling can considerably influence the MWI threshold \cite{jp4b-1mhy,bian:ipac2026-thp5653}. Bunch length variation can modulate the amplitude of the bunch wake and the exchange of head and tail part of the beam can alleviate the effect of the wake. In addition, transverse-longitudinal coupling can make particles with different betatron actions behave differently. Based on the classical formula, we have also provided a more general threshold evaluation method for the free space CSR induced MWI considering the bunch length variation and head-tail exchange around the ring.

\begin{figure}[htb]
	\includegraphics[width=8.5cm]{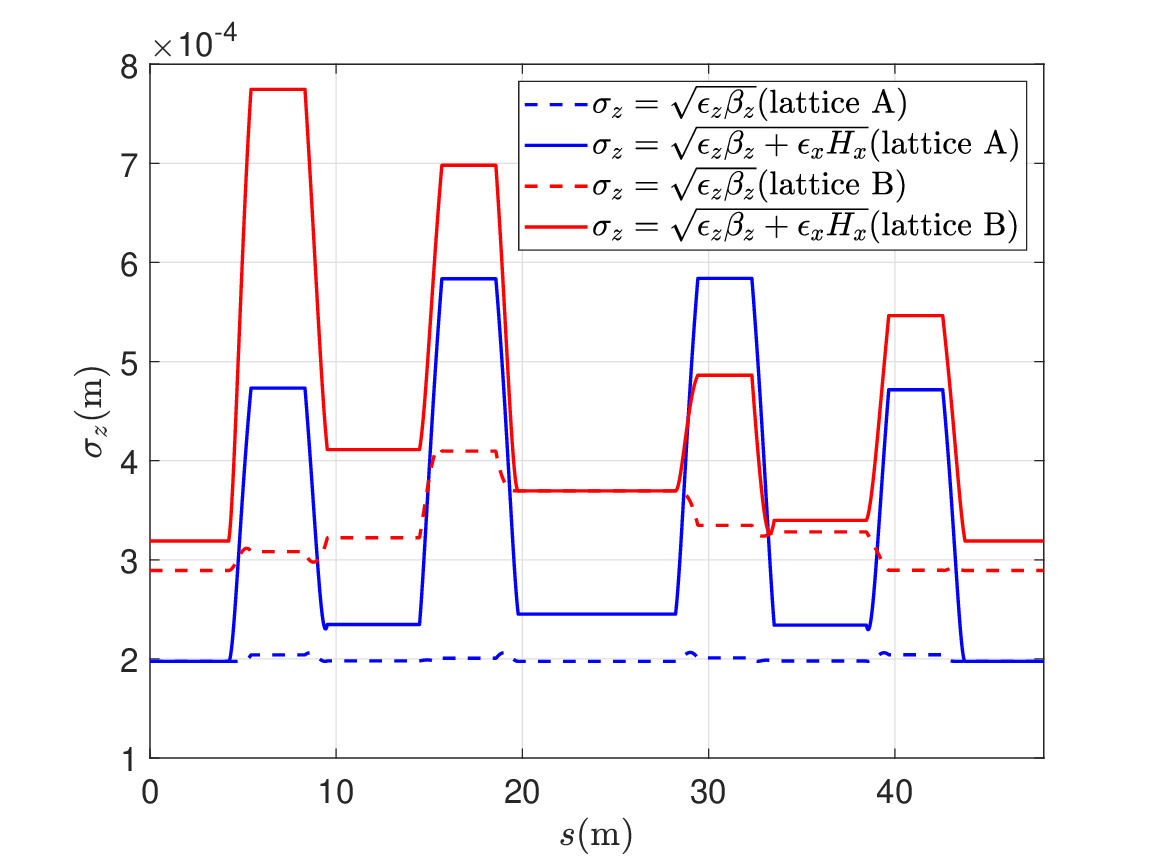}
	\caption{\label{fig:1} Bunch length evolution around the ring for the two lattices used in the experiment.}
\end{figure}

In this paper, we report the first experimental verification of circumferential bunch length variation and head-tail exchange affecting MWI in a storage ring with extremely small global phase slippage. The experiment was conducted at the Metrology Light Source (MLS) of the Physikalisch-Technische Bundesanstalt (PTB) in Berlin \cite{PhysRevSTAB.14.030705}. MLS is an electron storage ring with individually independent magnet power supplies, which has great flexibility in adjusting the lattice optics to achieve dedicated quasi-isochronous design. The bunch length evolution around the ring for the two lattices used in our experiment at the minimum achievable global phase slippage is shown in Fig.~\ref{fig:1}. These two quasi-isochronous lattices are named lattice A and B, respectively. Lattice A is a standard quasi-isochronous lattice with both small global and partial phase slippage, whereas lattice B has a much larger partial phase slippage. Other related parameters of the two lattices are shown in Table~\ref{tab:1}.

\begin{table}[b]
	\caption{\label{tab:1}%
		Parameters of the two lattices used in the experiment.
	}
	\begin{ruledtabular}
		\begin{tabular}{lc}
			Parameter & Value  \\
			\colrule
			Ring circumference & 48 m  \\
			Beam energy & 629 MeV  \\
		    Bending radius & 1.53 m \\
			RF voltage of lattice A  & 480 kV  \\
	 		Global phase slippage of lattice A & $3.2 \times 10^{-5}$ \\
			Energy spread of lattice A & $4.5 \times 10^{-4}$  \\
			Horizontal emittance of lattice A & 196.8 nm  \\
			RF voltage of lattice B  & 320 kV  \\
            Global phase slippage of lattice B & $3.6 \times 10^{-5}$ \\
			Energy spread of lattice B & $5.1 \times 10^{-4}$  \\
			Horizontal emittance of lattice B & 219.2 nm  \\
			
		\end{tabular}
	\end{ruledtabular}
\end{table}

Before presenting the experimental work, we briefly outline our theoretical analysis of the MWI threshold considering the bunch length variation and head-tail exchange around the ring. More detailed theoretical derivations can be found in our previous publications \cite{jp4b-1mhy,bian:ipac2026-thp5653}. In conventional storage rings, the bunch length remains constant within one turn, and the synchrotron tune $\nu_s$ is usually sufficiently small. Thus, the longitudinal relative positions of the particles in a bunch also remain approximately unchanged within one turn and the one-turn bunch wake is only related to the longitudinal position coordinate. For a Gaussian bunch, the one-turn bunch wake produced by free space CSR can be expressed as
\begin{eqnarray}
		W(z) &=& \frac{2^{5/6} \rho^{1/3}}{3^{7/3} \epsilon_0 \sigma_z^{4/3}}
		\left[
		\frac{\sqrt{\pi}q {}_1F_1\left(\frac{7}{6},\frac{3}{2},-\frac{q^2}{2}\right)}{\sqrt{3}\Gamma\left(\frac{5}{3}\right)} \right. \nonumber \\
	& &\left.	- \frac{2^{5/6}\Gamma\left(\frac{2}{3}\right) {}_1F_1\left(\frac{2}{3},\frac{1}{2},-\frac{q^2}{2}\right)}{\Gamma\left(\frac{7}{3}\right)}
		\right],
	\label{eq:3}
\end{eqnarray}
where $q=z/\sigma_z$, $\epsilon_0$ is the vacuum permittivity and ${}_{1}F_1$  is a confluent hypergeometric function of the first kind \cite{PhysRevAccelBeams.23.014402}. We found that the threshold of the free space CSR induced MWI
\begin{equation}
	N_{\text{th}} = \frac{\pi \nu_{s} \gamma \sigma_\delta \sigma_{z}^{4/3}}{r_{e} \rho^{1/3}}
	\label{eq:4}
\end{equation}
is inversely proportional to the standard divergence of the relative energy deviation change caused by the one-turn bunch wake, referred to as kick divergence
\begin{eqnarray}
	K &=& \frac{4\pi\epsilon_0 N r_e}{\gamma}
	\sqrt{\int \left(W(z)-\overline{W}\right)^2 f(z)dz} \nonumber \\
	&=& 1.545\frac{N r_e \rho^{1/3}}{\gamma \sigma_z^{4/3}},
	\label{eq:5}
\end{eqnarray}
where $N$ represents the number of particles per bunch, $\sigma_\delta$ is the natural energy spread, $r_e$ is the classical electron radius, $\overline{W}=\int W(z)f(z)dz$, and $f$ represents the density distribution of the bunch. Note that $\rho$ and $\sigma_z$ are the two parameters that directly related to the strength of the CSR wake.

When the bunch length evolution is included, a direct impact is that the amplitude of the one-turn bunch wake changes accordingly owing to the dependence of the wake on longitudinal relative distance. Nevertheless, the wake perturbation accumulated over one turn remains weak, so we can still use one-turn bunch wake to determine the instability threshold. The corresponding one-turn bunch wake can be expressed as
\begin{eqnarray}
	&&W^{\text{n}}(x,x',z,\delta) = \nonumber \\
	&&\int_{\text{dipole}}
	\frac{W\left(z + R_{56}(s)\delta + R_{51}(s)x + R_{52}(s)x'\right)}{2\pi\rho}ds.
	\label{eq:6}
\end{eqnarray}
Note that the wake depends not only on longitudinal position, but also on relative energy deviation and horizontal coordinates. Classical MWI analysis is purely longitudinal, resting on the implicit assumption that particles with different transverse coordinates experience identical longitudinal dynamics. Inclusion of transverse-longitudinal coupling breaks this degeneracy, leading to the one-turn longitudinal bunch wakes differ for particles with distinct transverse coordinates. Owing to the betatron oscillation, a reasonable approach is to average the one-turn longitudinal bunch wake according to the betatron action, and then determine the instability thresholds separately for particles with different betatron actions. For particles with each betatron action $J_x$, we calculate the  kick divergence $K^{\text{n}}(J_x)$ and the bunch wake rotation angle $R(J_x)$ characterizing the effect of the head-tail exchange in the longitudinal phase space. Then the threshold can be expressed as
\begin{equation}
	N^{\text{n}}_{\text{th}}(J_x) = \left(1 - D(J_x)\right)\frac{K}{K^{\text{n}}(J_x)}N_{\text{th}},
	\label{eq:7}
\end{equation}
where $D(J_x)=0.89e^{-0.44R(J_x)}+0.11e^{2.31R(J_x)}-1$ is a fitting formula, $K$ and $N_{\text{th}}$ denote the kick divergence and threshold obtained without considering the bunch length variation and head-tail exchange, respectively. Finally, we define the global threshold of the bunch as the bunch population at which all particles with betatron action no greater than 3$\epsilon_x$ experience instability, which is given by
\begin{equation}
	\widetilde{N_{\text{th}}} = \max{ N^{\text{n}}_{\text{th}}(J_x \leq 3\epsilon_x)}.
	\label{eq:8}
\end{equation}

\begin{figure}[htb]
	\includegraphics[width=8.5cm]{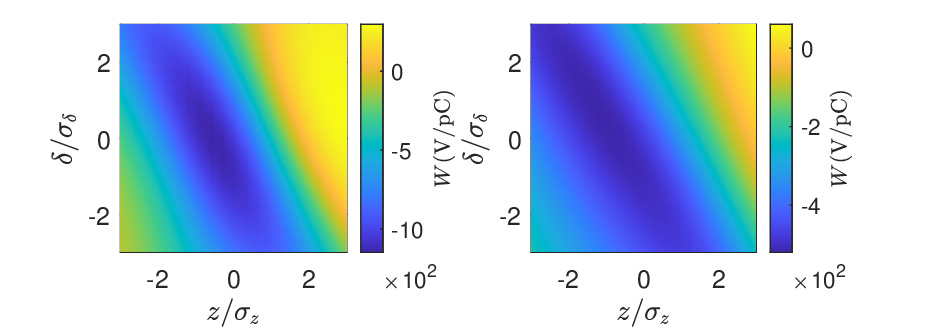}
	\caption{\label{fig:2} One-turn longitudinal bunch wake in longitudinal phase space at horizontal actions of 0 (left) and $3 \epsilon_x$ (right).}
\end{figure}

\begin{figure}[htb]
	\includegraphics[width=7.5cm]{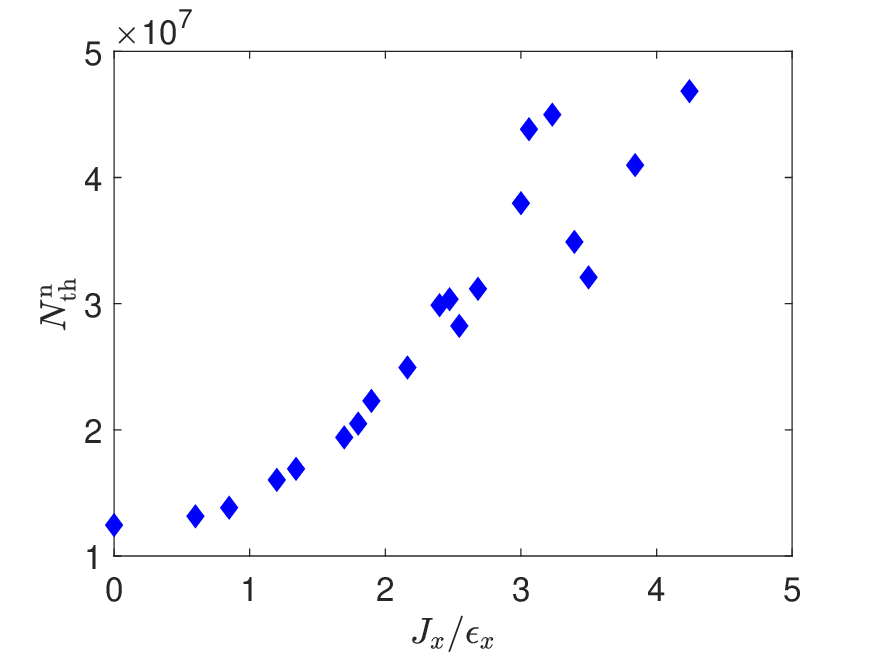}
	\caption{\label{fig:3} Thresholds for particles with different horizontal actions.}
\end{figure}

\begin{figure}[htb]
	\includegraphics[width=7.5cm]{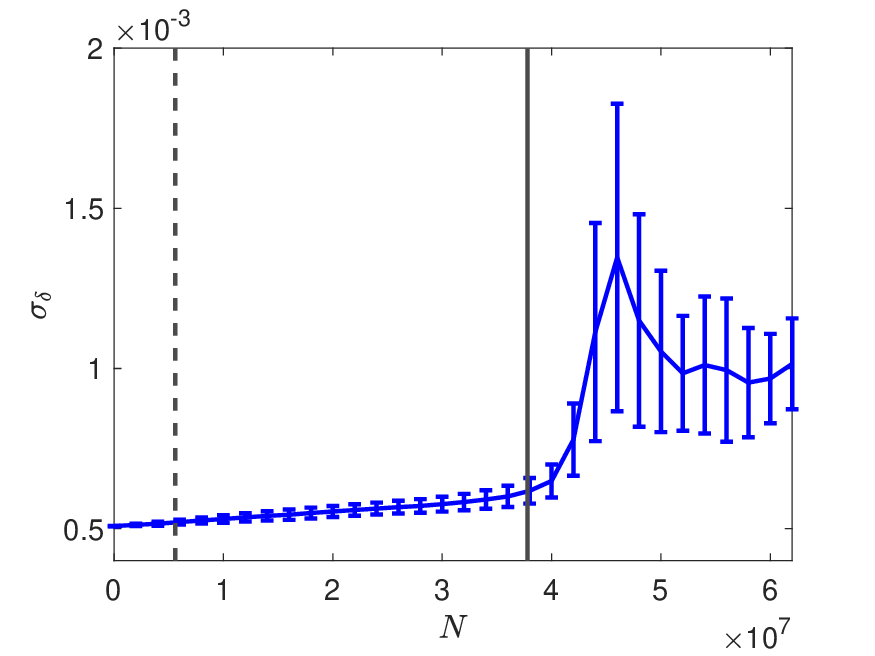}
	\caption{\label{fig:4} Energy spread as a function of bunch population. The black dashed line represents the classical prediction while the black solid line represents the threshold predicted by our new method.}
\end{figure}

Next, we would like to use lattice B employed in our experiment at the minimum achievable global phase slippage as a specific example to predict its threshold based on our theory. Figure~\ref{fig:2} presents the one-turn longitudinal bunch wake in longitudinal phase space at horizontal actions of 0 and $3 \epsilon_x$, which are calculated via Eq.~(\ref{eq:6}) and averaged, respectively. Significant discrepancies can be observed, particularly in amplitude, which is consistent with our theoretical expectation. Then we get the thresholds for particles with different horizontal actions, as illustrated in Fig.~\ref{fig:3}. The calculated results indicate that particles with smaller horizontal action have lower instability threshold in this example. Accordingly, we take $\widetilde{N_{\text{th}}}=3.78\times10^7$ as the threshold of the entire bunch. We have also conducted particle tracking simulations. The significant increase in the energy spread of the bunch relative to the natural energy spread is usually used as the sign of instability occurrence in the MWI simulation. The simulation results are shown in Fig.~\ref{fig:4} and it can be seen that the simulated threshold aligns well with our prediction.

Now, we present our experimental work. The minimum achievable global phase slippage in the experiment is $3.2 \times 10^{-5}$ for lattice A and $3.6 \times 10^{-5}$ for lattice B. The bunch length evolution around the ring for lattice A and B at the minimum global phase slippage is shown in Fig.~\ref{fig:1} and the related parameters are shown in Table~\ref{tab:1}. Under this setting, the partial phase slippage and transverse-longitudinal coupling can already make bunch length vary around the ring. We can also tune the global phase slippage by slightly changing the quadrupole currents while keeping the dispersion function pattern unchanged.

The essential feature of the MWI is the formation of microstructures arising from density modulation in longitudinal phase space. However, at the instability threshold the density modulation is small and can hardly be detected through conventional measurements of bunch length or energy spread in the experiment. Therefore, we use THz radiation generated by the microstructures in the bunch to determine the threshold, which is more sensitive and is the common method to measure MWI threshold in the experiment. In our experiment, a THz sensitive Schottky barrier diode detector from ACST \cite{acst_website} is used to detect the THz signal and a frequency spectrum analyzer is connected for subsequent data processing. Benefiting from the flexibility of the MLS control system, the measurement process can be automated. The algorithm can be sketched as: set a beam current, record Fourier transform data of the THz signal across a series of global phase slippage, use automatic scraping procedure to reduce the beam current by 5\%, and repeat. All measurements are performed in single bunch operation to avoid the influence of other effects such as multi-bunch effect.

\begin{figure}[htb]
	\includegraphics[width=7.5cm]{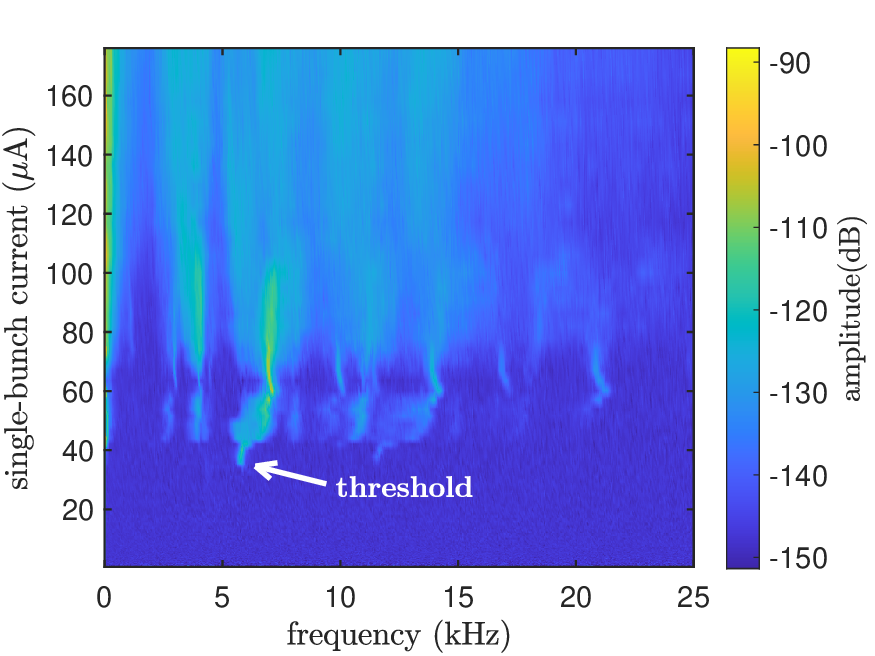}
	\caption{\label{fig:5} Spectrogram of the THz signal as a function of the single-bunch current for a synchrotron frequency of 3 kHz with the applied RF voltage 320 kV.}
\end{figure}

\begin{figure}[htb]
	\includegraphics[width=8cm]{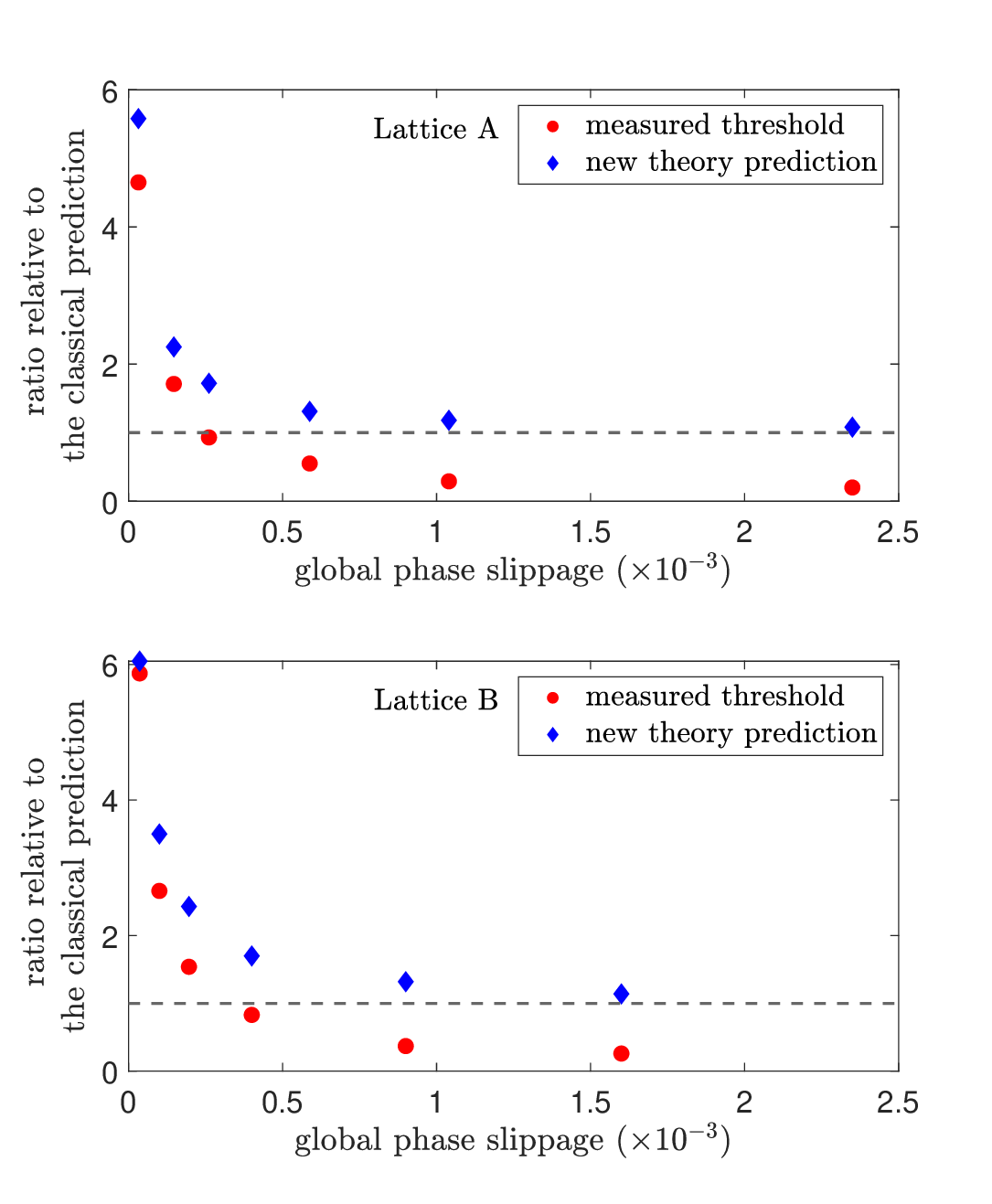}
	\caption{\label{fig:6} Experiment measurement results at different global phase slippage and comparison with theory for lattice A (top) and lattice B (bottom). Red circles mean the ratio of the measured threshold to the classical prediction and blue diamonds mean the ratio of the threshold derived from our new theory to the classical prediction.}
\end{figure}

Figure~\ref{fig:5} shows an example spectrogram of the THz signal as a function of the single-bunch current for lattice B at the minimum global phase slippage, which corresponds to a synchrotron frequency of 3 kHz with the applied RF voltage 320 kV. It can be seen that distinct THz signals appear at synchrotron frequency and its harmonics as the current increases and the spectrum becomes broadband at higher current. Thus, the measured instability threshold current is about 34.6 $\mu \mathrm{A}$, corresponding to the number of electrons in the bunch being $N = 3.46 \times 10^7$. Under these parameters, the wake in the storage ring is dominated by free space CSR, and the shielding effect can be neglected. The measured result is close to our theoretical prediction, and is approximately six times that predicted by the classical formula. Figure~\ref{fig:6} presents the experiment measurement results at different global phase slippage and comparison with theory for lattice A and B, respectively. It is observed that the ratio decreases gradually with the increase of the global phase slippage for both lattices. The trend of experimental results agrees reasonably well with our new theory. Theoretically, the ratio should approach 1 when the global phase slippage is large enough, since the storage ring is similar to a conventional classical storage ring. We attribute the observed deviation to the effects of other wakes when the bunch length is large.



In conclusion, we state that our experimental work supports our theoretical analysis, indicating that circumferential bunch length variation and head-tail exchange can increase the threshold of MWI. Our results have significant implications for the design of short bunch storage rings, demonstrating that the instability threshold can be increased by lattice optics optimization or even actively increasing the transverse emittance. Although our work focuses on the free space CSR wake, the underlying physical mechanism is general and applicable for other types of wake.

\clearpage

\begin{acknowledgments}
This work is supported by the National Natural Science Foundation of China (NSFC Grants No. 12035010 and No. 12522512) and the National Key Research and Development Program of China (Grant No. 2022YFA1603401).

\end{acknowledgments}


\appendix

\nocite{*}


\bibliography{apssamp}

\end{document}